\documentstyle[11pt,aas2pp4,psfig]{article}

\begin{document}
\title{ Induced Raman Scattering in Pulsar Magnetospheres}
\author { Maxim Lyutikov}
\affil{
Theoretical Astrophysics, California Institute of Technology,
Pasadena, California 91125, USA {\footnote {Currently at the
Canadian Institute for Theoretical Astrophysics, 
60 St George Street, Toronto, M5S 3H8} }  } 

\setcounter{footnote}{0}

%\date { }

\begin{abstract}
It is shown that 
induced Raman scattering of electromagnetic
waves in the strongly magnetized electron-positron
plasma of pulsar magnetosphere may be  important for  wave
propagation and as an effective saturation mechanism for  electromagnetic
instabilities. 
The frequencies, at which strong Raman scattering occurs
in the outer parts of magnetosphere, fall into the observed
radio band.
The typical threshold
intensities for the strong Raman scattering are of the order
 of the observed intensities, implying that pulsar magnetosphere
may be optically thick to Raman scattering of electromagnetic waves.

\end{abstract}

%\begin{keywords}

%\end{keywords}

\onecolumn

\section*{}

\onecolumn 
\setcounter{page} {2}
\setcounter{section} {0}

\onecolumn

\section{Introduction}

We aim to estimate the effects of the the induced Raman
scattering of the strong electromagnetic wave propagating 
in pulsar magnetosphere. Raman scattering may be considered as 
a parametric decay of the initial transverse electromagnetic wave
in the another electromagnetic wave and plasma wave. The probability
of this process is greatly enhanced if there many waves present in the 
final states. This is induced Raman scattering.
Another type of induced scattering -  induced Brullion scattering,
i.e. the decay of the initial transverse electromagnetic wave
in the another electromagnetic wave and ion sound is prohibited
in electron-positron plasma, since pair plasma does not support
low frequency, density perturbing waves (like ion sound wave
in electron-ion plasma).

Strong nonlinear coupling occurs when two waves beat together 
and the sum or difference frequency and wavelength match the 
frequency and wavelength of the third wave:
\begin{eqnarray}
&& 
\omega_3= \omega_1+ \omega_2 
\mbox{} \nonumber \\ \mbox{}
&&
 k_{z,3}=  k_{z,1}+  k_{z,2}
\label{n}
\end{eqnarray}
 In quantum language these relations may be interpreted as
conservation of energy and momentum along the magnetic field, respectively.
(the transverse components of the momentum are not conserved).
The probability of the  induced scattering depends on the number of waves
in the final state, which, in turn, depends on the damping  or escape rate of the 
daughter waves. 
 The energy
transfer between the modes will be efficient if the 
energy of the pump wave is strong enough to overcome the 
damping losses or escape  of the generated waves.
Thus, the induced Raman
scattering is a threshold process: if the intensity of the 
pump wave exceeds the threshold value, the initial
electromagnetic wave would start converting energy into the 
decay waves decreasing its amplitude exponentially.

Induced Raman
scattering may be important in the pulsar environment in
two ways. First, it may provide en effective damping
of the existing electromagnetic wave, that has been generated 
by some emission mechanism at the lower altitude in the 
pulsar magnetosphere, where the resonant conditions for the 
induced Raman
scattering  were not satisfied. This may result 
in a short time variability which is generally 
observed in pulsar radio emission. Secondly,
it may provide an effective saturation mechanism for the
growth of the electromagnetic wave provided  that the conditions
for the wave excitation by some mechanism are satisfied
in the region where an effective Raman
scattering takes place.

The first possibility, i.e. the scattering of the existent 
wave, is  simpler to consider. As a first approximation,
we can treat the intensity of the pump wave as a constant. 
Then the nonlinear equations describing the wave coupling
become linear in amplitudes of the decayed waves. The 
exponentially growing solutions will imply an effective
energy transfer from the pump wave.
As the intensities of the decay waves grow, 
this approximation breaks down in two cases: when the amplitudes
of the decay waves become comparable to the pump wave
or when the amplitude of the decay waves enter nonlinear
stage and  the waves   start loosing energy due to some nonlinear process
(like particle trapping and acceleration).
The net effect of any energy loss by the decay wave is the depletion
of the original pump wave. For the practical estimates,
we can neglect the nonlinear stages of evolution 
(like cyclic energy transfer between the pump and the daughter waves)
and assume that if the intensity of the pump wave begins to
decrease exponentially in the linear stage, then the wave decays
completely.

The second possibility, e.g. when the induced Raman
scattering  provides a nonlinear saturation mechanism, is more complicated,
since the intensities of all waves may be of the same order. 
A considerable simplification in this case may be obtained if the 
damping of one of the decay  modes is very strong or if it leaves the
region of the resonant interaction fast enough. Then, 
after a short period of time (when the intensities of the other
weakly damped waves grow considerably) the intensity of this damped
 mode is much
smaller than the intensities of the two other modes and can
be neglected.

In what follows we neglect the possible nonlinear stages of the
development of Langmuir turbulence (particle trapping or quasilinear
diffusion). We also assume that the pump wave is broadband. This
is quite different from the conventional laboratory case of a monochromatic
laser-plasma interaction. The condition of a broadband pump wave
implies that its band width $\Delta \omega$ is  much larger than the typical
growth rate of the decay instability $\Gamma$:
\begin{equation}
\Delta \omega \gg \Gamma
 \label{kd941}
\end{equation}
If this condition is satisfied, then we can use a random phase approximation
for the statistical description of the interacting waves.
On the other hand, the condition of weak turbulence allows one to calculate
the 
matrix elements of the interaction in the approximation of stationary
phases. 

The general expression for the third order nonlinear current
in plasma in a magnetic field
has been written down by \cite{TsytovichShvartsburg}. An extremely
complicated form of the corresponding expressions 
makes the general case of Raman scattering very difficult
to consider. Several important simplifications can be done when
considering induced Raman scattering  in the pulsar magnetosphere.
First, the superstrong magnetic field allows an expansion of the
currents in $1/\omega_B$. Second, 
in the  pair plasma with the same distributions
of electrons and positrons some of the  nonlinear 
currents cancel out since they are proportional to the third 
power of the electric charge (this
cancellation
is exact in the unmagnetized electron-positron plasma).
 The third, less justified approximation, is 
that we will make is that all the three interacting waves
propagate along magnetic field. This is an important assumption.
It allows us to simplify the consideration
considerably and to obtain some analytical estimates of the effects
of induced Raman scattering. 

 A short overview of the work on the nonlinear process in the 
pulsar magnetosphere will be appropriate here.
The possible decay  processes  for the transverse waves
and the corresponding references  are
($t$ denotes transverse wave, $l$ denotes longitudinal wave and $e$ denotes
a charged particle):
(i)  a decay of a transverse wave into two
transverse wave $ t \rightarrow t ^{\prime} +
t ^{\prime \prime}$
(\cite{GedalinMachabeliPismaAj}),
(ii) a decay of a transverse wave into two Lagmiur waves
$ t \rightarrow l+l^{\prime}$
(\cite{GedalinMachabeli1983}),
 (iii) a decay of a transverse wave into
another transverse and Lagmuir wave
 $ t \rightarrow t ^{\prime} + l$
(\cite{GedalinMachabeli1983}) \footnote{ The final answer for the 
matrix element contains probably an insignificant  typographical error}
(iv) induced scattering of transverse waves 
$ t + e \rightarrow t ^{\prime} + e $ 
(\cite{BlandfordScharleman}, \cite{SincellKrolik1992}, \cite{OchelkovUsov},
\cite{Wilson82}).
The possible   processes  for Lagmuir waves are
(i)   a decay of a Lagmuir wave into another Lagmuir
and transverse wave $  l \rightarrow t  + l ^{\prime}$
(\cite{GedalinMachabeli1983}), (ii)
a decay of a Lagmuir wave into two transverse waves
$  l \rightarrow t  +
t ^{\prime}$
(\cite{Mikhailovski1980}),
(iii) Lagmuir wave merger
 $  l +  l ^{\prime} \rightarrow t$
(\cite{MachabeliMamradzeMelikidze1982}),
(iii) induced scattering of  Lagmuir waves
(\cite{Lyubarskii93}).
When treating the nonlinear process involving 
transverse waves, the matrix elements in above works have
been calculated in the drift approximation -  the  expansion  in
parameter $1/\omega_B$ was done in the very beginning ($\omega_B=
|q| B/m c$ is the positive nonrelativistic cyclotron frequency, 
$B$ is magnetic field, $m$ is mass of an electron and $c$ is the
speed of light).

\section{ Kinematics of Raman Scattering in Pair Plasma}
\label{Kinematics}

\subsection{Wave dispersion}
\label{Wavedispersion}

Normal modes of relativistic 
pair plasma for the case of parallel propagation consist of two
transverse waves with the dispersion relation 
\begin{equation}
\, \omega = k c  ( 1- \delta), \hskip .1 truein
\delta = {\, \omega_p^2 T_p \over  \omega_B^2 }
\label{det33}
\end{equation}
and a plasma Langmuir wave  with a dispersion relation
\begin{equation}
\omega^2= { 2\, \omega_p^2 \over \, T_p }+ k^2 c^2 \beta_T^2
\label{lliafs3}
\end{equation}
where in Eq. (\ref{lliafs3}) the wave  vectors should not be much larger than
cross-over point:
\begin{equation}
\omega_o^2 = k_o^2 c^2 =
2 \omega_p^2 < \gamma ( 1+\beta )^2 >
\label{n15}
\end{equation}
(\cite{LyutikovPhD}).
We have introduced here a plasma frequency in the plasma frame
$\omega_p^2 = 2 \omega_B \Omega \lambda/\gamma_p$, where $\Omega$
 is the pulsar rotation angular velocity, $\lambda$ is the
multiplicity factor, $\gamma_p$ is the Lorentz factor of the plasma frame,
$T_p \approx <\gamma> $ is Lorentz invariant temperature of plasma,
$\gamma$ is Lorentz factor of a particle and $\beta_T =
\sqrt{ T_p^2-1}/T_p$ is a characteristic (dimensionless)
 thermal velocity of particles, $\beta$ is (dimensionless) velocity.
The brackets denote averaging over the distribution function.

At the point $\{ \omega_o,k_o\}$ the dispersion curve of plasma
waves intersect that of waves in vacuum, so that for $k > k_o$
the plasma waves become subluminous. The necessary requirement
for that is that the distribution function falls fast enough at large
values of momentum (faster than $1/p^4$) (\cite{LominadzeMikhailovskii}).

In the subluminous region, the plasma
wave dispersion relation may be written as
\begin{eqnarray}
&&
\omega = kc - \eta ( k - k_o) c, \hskip .2 truein \mbox {where}
\mbox{} \nonumber \\ \mbox{} &&
\eta = 1-  \left({\partial \epsilon _{zz} \over \partial k} \right) \left/
 \left(
{\partial \epsilon _{zz} \over \partial \omega } \right) \right|_
{\omega_o, k_o} =
{ < \gamma ( 1+ \beta)^2> \over < \gamma^3 ( 1+\beta)^3> } \,
\approx {1\over T_p^2 }
\label{n5}
\end{eqnarray}

\subsection{Scattered frequencies}
\label{Scatteredfrequencies}

From the resonant conditions
\begin{eqnarray}
&&
\omega_3= \omega_1+ \omega_2 
\mbox{} \nonumber \\ \mbox{}
&&
k_3=k_1+k_2 
\label{r61}
\end{eqnarray}
it follows that that the only kinematically allowed
Raman scattering  process is a back scattering ($k_1 < 0$) of the initial electromagnetic
wave (see Fig. 1).
%ref{RamanFig}). 
We find  then 
\begin{eqnarray}
&& k_3 ={1\over 2 c } \left( {\omega_2 \over 1-\delta} + k_2 \, c\right)
 \mbox{} \nonumber \\ \mbox{}
&&
|k_1| = {1\over 2 c } \left(-  {\omega_2 \over 1-\delta} +k_2 \, c\right)
\label{r62}
\end{eqnarray}
 In the region just above the cross-over point $\{ \omega_o, k_o\}$
 it is possible to use 
the expansion (\ref{n5}) for the dispersion of the plasma waves.  
The resonant wave vectors are then given by
\begin{equation}
k_3 = { k_2+k_o\over 2}  \left( 1 - { k_2 - k_o \over 4 T^2 k_o  } \right),
\hskip .3 truein 
|k_1|= { k_2 - k_o \over 2  } 
\label{r63}
\end{equation}
The wave vectors in Eq. (\ref{r63}) are limited by $ k_2 -k_o \leq k_o$, $ 
|k_1| \leq k_o  $.

\subsection{ Induced Raman Versus Induced Compton Scattering}
\label{RamanCompton}

Induced Raman scattering is closely related
to the  induced Compton scattering of waves by the plasma
particles. In the conventional treatment of induced Compton scattering
in pulsar setting
(e.g. \cite{BlandfordScharleman}, \cite{SincellKrolik1992})
the collective effects of the plasma are ignored. 
In the case of induced  Raman scattering collective effects 
play a major role. Raman scattering is a limit of the 
induced Compton scattering of waves by particles in a medium when
the beat wave of the incoming and scattered wave becomes a normal
mode of the medium (equivalently, it falls on the "mass surface" 
of the medium dispersion equation).

In the nonrelativistic plasma 
the collective treatment of the wave scattering by plasma particles
 is 
justified if the following  
 condition is  true:
\begin{equation}
\Delta k \lambda_D \ll 1 
\label{n3} 
\end{equation}
where $\Delta k$ is the change in the
wave number of the scattered wave and $\lambda_D $ is the Debye length.
 Condition (\ref{n3})  states that the
wave number of the oscillations of the electron on the beat of the
two scattered waves be much less than the inverse of the  Debye length.
If condition (\ref{n3}) is not satisfied, then the beat wave will either
experience a strong Cherenkov damping ( if $ \Delta k \lambda_D
\approx  1$) or
the beat wave will not feel the presence of a medium
( if $ \Delta k \lambda_D \gg 1$), so that
the scattering process will be described as induced Compton scattering.

In the relativistic plasma with the characteristic
Lorentz factor  of the particles
$ < \gamma> \,\gg 1$  the condition 
(\ref{n3}) is modified.
To find the the generalization of condition (\ref{n3}) to the
case of relativistic plasma we note, that 
condition (\ref{n3})  may be rewritten in therm of the 
phase velocities of the beat wave $v_{ph} = \omega/k$
 (which is a plasma wave in case
of Raman scattering) and the thermal velocities of the plasma particles 
$v_T = < v ^2> ^{1/2} $. Condition (\ref{n3}) then can be rewritten
 \begin{equation}
v_{ph} \gg v_T 
\label{n31}
\end{equation}

Using (\ref{n5}), we find that the phase velocity of the subluminous 
plasma waves becomes equal to the thermal velocity of plasma
particles approximately at $ k \approx 2  k_o$. Thus,
the following condition should hold for the 
wave-particles scattering in relativistic pair 
plasma to be considered as a Raman scattering:
\begin{equation}
\Delta  k < 2  k_o \approx 4 \sqrt{2 < \gamma>} \omega_p
\label{n6}
\end{equation}
When $\Delta  k \approx 2  k_o$ the plasma waves are strongly 
damped and when  $\Delta  k \gg 2  k_o$ the wave-particles scattering 
is described as single particle Compton scattering.
As we will see later (Eq. (\ref{ttt1})) the condition
(\ref{n6}) is satisfied in our problem.

\section{Transition probabilities}
\label{Transitionprobabilites}

First we calculate the matrix elements for the nonlinear
three wave interaction in the strongly magnetized
electron-positron plasma when all three waves have their wave
vectors along the magnetic field. 
We represent the electric fields of the the three interacting waves
in the form
\begin{equation}
{\bf E} ( {\bf r }, t)=  \Re \left[
{\bf E}_1 \exp \{ i( {\bf k}_1 {\bf r } - \omega_1 t ) \}+
{\bf E}_2 \exp \{ i( {\bf k}_2 {\bf r } - \omega_2 t ) \}+
{\bf E}_3 \exp \{ i( {\bf k}_3  {\bf r } - \omega_3 t ) \} \right]
\label{n1}
\end{equation}
The indices $1,2,3$ refer to the scattered electromagnetic wave,
plasma wave and the initial electromagnetic wave  respectively, 
$\Re $ denotes the real part.
We consider lineally polarized waves. For the parallel propagation
the case of circularly polarized waves may be trivially reduced
to the scattering of two lineally polarized waves.
Implicit assumption here is that the coupling between the waves
is weak, so that the waves retain their identity as eigenmodes
of the medium during the times much longer than inverse of their
frequencies.

We calculate transition probabilities by finding the  coupling
coefficients for the resonant three-wave interaction. 
The interaction of two transverse and one longitudinal wave 
propagating along magnetic field  in a 
relativistic plasma has been considered by \cite{BrodinStenflo}.
For the decay process $ \omega_3 = \omega_1+ \omega_2$ and
$ k_3= k_1+ k_2$ the coupling of the waves is described by the
system of equations
\begin{eqnarray}
&&
{ d E_1 \over dt} = c_1  E_2^{\ast} \, E_3
\mbox{} \nonumber \\ \mbox{}
&&
{ d E_2 \over dt} = c_2  E_1^{\ast} \, E_3
\mbox{} \nonumber \\ \mbox{}
&&
{ d E_3  \over dt} = c_3  E_1^{\ast} \, E_3
\label{r}
\end{eqnarray}
where
\begin{eqnarray}
&&
 c_1 = - { \omega_1^2 \over \left( { \partial  \omega^2 \epsilon^{\perp}_{\pm}
 \over  \partial  \omega}
\right) _1 } C_{\pm}
\mbox{} \nonumber \\ \mbox{}
&&
c_2 = - {  1 \over  \left( { \partial  \epsilon_{zz} 
 \over  \partial \omega}
\right)  _2} \, C_{\pm}
\mbox{} \nonumber \\ \mbox{}
&&
c_3= { \omega_3^2 
 \over \left( { \partial  \omega^2 \epsilon^{\perp}_{\pm}
 \over  \partial \omega} \right) _3} \, C_{\pm}
\label{r1}
\end{eqnarray}

and 
\begin{eqnarray}
&&
 \epsilon_{zz} = 1+ \sum {\omega_p^ 2 \, m \over k} \,
\int { d {\bf p} \over \omega - k v_z } \,
 { \partial F ( {\bf p}) \over \partial p_z}
\mbox{} \nonumber \\ \mbox{}
&&
 \epsilon^{\perp}_{\pm}=
1+ {1\over 2} \, \sum  { \omega_p^ 2 \over \omega^ 2} \,
\int { d {\bf p} \over \gamma} \, \left[ 
{ ( \omega - k v_z) p_{\perp} \over
\omega - k v_z \mp \omega_c}{ \partial F ( {\bf p}) \over \partial p_{\perp} }
\,+ { k \,  p_{\perp}^2 \over 
m \gamma (\omega - k v_z \mp \omega_c) }
  { \partial F ( {\bf p}) \over \partial p_z}
\right]
\mbox{} \nonumber \\ \mbox{}
&&
 C_{\pm} = -  \sum {\omega_p^ 2 q \over \omega_1 \omega_2 m} \int d {\bf p} \,
 \left\{ { m\over \gamma \hat{ \omega}_2 } 
 { \partial F ( {\bf p}) \over \partial p_z}
\right.
\mbox{} \nonumber \\ \mbox{}
&&
\mp 
{ \partial  \over \partial p_{\perp}^2 }
\left[ {\omega_c  p_{\perp}^2 \over  \gamma^4 \hat{ \omega_2}} 
\left( { k_3 \over (\hat{ \omega}_1 \mp \omega_c) \,
 (\hat{ \omega}_3 \mp \omega_c) } -
{ k_1  \over (\hat{ \omega}_1 \mp \omega_c) (\hat{ \omega}_3 \mp \omega_c) } -
{\gamma^2 v_z \over c^2} {  \hat{ \omega}_1 + \hat{ \omega}_3 \mp \omega_c 
\over \omega_1 \omega_3 } \right) \right] F ( {\bf p}) 
\mbox{} \nonumber \\ \mbox{}
&&
\left.
- {   p_{\perp}^2 \over 2  \gamma^3  m c^2 \hat{ \omega}_2 } 
\left[ 1- { k_1 k_3 c^2 \over  \gamma^2 (\hat{ \omega}_1 \mp \omega_c) \,
 (\hat{ \omega}_3 \mp \omega_c) } - \right. \right.
\mbox{} \nonumber \\ \mbox{}
&&
\left. \left.
v_z \left( {k_3 \over  (\hat{ \omega}_3 \mp \omega_c)} +
{k_1  \over  (\hat{ \omega}_1 \mp \omega_c)} \mp
{\omega_c k_1 \over  (\hat{ \omega}_1 \mp \omega_c) \,
 \hat{ \omega}_2  } \pm
{\omega_c k_3  \over  (\hat{ \omega}_3 \mp \omega_c) \,
 \hat{ \omega}_2  } \right)  \right]  
\right.
\left.  
\times 
 { \partial F ( {\bf p}) \over \partial p_z}
\right\}
\label{r3}
\end{eqnarray}
where $ \hat{ \omega} _{1,2,3} =  \omega _{1,2,3} - 
k _{1,2,3} v_z$ and $ \omega_c =  q B_0/ \gamma m c$ is the relativistic 
cyclotron frequency.

We now simplify this relations for the case of one dimensional pair
plasma with the same distribution functions. Setting
$ F  ( {\bf p}) = {1\over \pi} \delta(  p_{\perp}^2 ) f(p_z) $ 
and introducing a positive nonrelativistic gyrofrequency
$ \omega_B = \sigma_q q B_0/  m c$ ($\sigma_q$ is the sign of the charge)
we find

\begin{eqnarray}
&&
 \epsilon_{zz} = 1+ { 2 \omega_p ^2 \over k} 
\int { dp_z \over \omega - k v_z } \,
 { \partial  f(p_z) \over \partial p_z}
\mbox{} 
\label{r401}
\\ \mbox{}
&&
 \epsilon^{\perp}=
1- {2 \omega_p ^2} \int { dp_z \over \gamma} 
{ (\omega - k v_z)^2 \over 
(\omega - k v_z)^2 - \omega_B^2/\gamma^2 }  f(p_z)
\mbox{} 
\label{r402}
 \\ \mbox{}
&&
C=- { 2 \omega_p ^2 |q| \over \omega_1  \omega_3 m} \int 
{ dp_z \over \gamma^5} {  \omega_B \over \hat{ \omega}_2 ^2} \,
\left( {k_3 \hat{ \omega}_1 \over
\hat{ \omega_1}^2 -  \omega_B^2/\gamma^2 } -{
k_1  \hat{ \omega}_3 \over
\hat{ \omega_3}^2 -  \omega_B^2/\gamma^2 } \right)  f(p_z)
\label{r4}
\end{eqnarray}

The probability of the decay is (\cite{Melrosebook1} Eq. 10.124)
\begin{eqnarray}
&&
u({\bf k}_3 ,{\bf  k_2},{\bf k}_1)  = 
 ( 2 \pi)^5 \hbar { R_E^1 R_E^2 R_E^3 \over 
\omega_3 \omega_1 \omega_2} \left|
 \omega_1 \omega_2 \omega_3  C \right|^2 
\delta({\bf k}_3 -{\bf  k_2}-{\bf k}_1) \, 
\delta(\omega_3 - \omega_2 - \omega_1)
= 
\mbox{} \nonumber \\ \mbox{}
&&
{ ( 2 \pi)^5  \hbar q^2 \omega_p ^4  \omega_B^2 \omega_2  \over 2  m^2 
 \omega_1 \omega_3 }  
\left| 
 \int { dp_z \over \gamma^5  \hat{ \omega}_2 ^2} \,
\left( {k_3 \hat{ \omega}_1 \over
\hat{ \omega_1}^2 -  \omega_B^2/\gamma^2 } -{
k_1  \hat{ \omega}_3 \over
\hat{ \omega_3}^2 -  \omega_B^2/\gamma^2 } \right)  f(p_z)
\right|^2
\mbox{} \nonumber \\ \mbox{}
&&
\delta({\bf k}_3 -{\bf  k_2}-{\bf k}_1) \, 
\delta(\omega_3 - \omega_2 - \omega_1)  
\label{r411}
\end{eqnarray}

with $  R_E^1 =  R_E^2 =  R_E^3 =1/2$

In the cold case we have

\begin{equation}
u({\bf k}_3 ,{\bf  k_2},{\bf k}_1)  =
{  ( 2 \pi)^5  \hbar q^2  \omega_p ^4  \omega_B^2  \over 2  m^2
\omega_1 \omega_3 \omega_2 ^3 }
\left|
 {k_3 \omega_1 \over  \omega_1 ^2 -  \omega_B^2  } -{
k_1  \omega_3  \over
\omega_3^2 -  \omega_B^2 } 
\right| ^2 
\delta({\bf k}_3 -{\bf  k_2}-{\bf k}_1) \, \delta(\omega_3 - \omega_2 - \omega_1)  
\label{r412}
\end{equation}
which may be obtained from \cite{Melrosebook1} Eq. 10.105.

Equations (\ref{r401}) and (\ref{r4})
can be simplified in the limit of a very strong magnetic
field. In the limit $\hat{\omega} \ll \omega_B/\gamma$ we find
\begin{eqnarray}
&& \epsilon^{\perp}=
1+  {2 \omega_p ^2 \over \omega_B^2} \int { dp_z}  \gamma 
(\omega - k v_z)^2   f(p_z)
\mbox{}  \\ \mbox{}
&&
C= \pm { 2 \omega_p ^2 |q| \over  m} \int
{ dp_z \over \gamma^3} { 1\over  \hat{ \omega}_2 ^2 \omega_B }
\left( {k_3 \over \omega_3 }- {k_1 \over \omega_1 }\right)
 f(p_z)
\label{r5}
\end{eqnarray}

Using Eq. (\ref{r401}) 
the equation  (\ref{r5}) may be rewritten in the form
\begin{equation}
C= \pm { 2  |q| \over m  \omega_B } 
\left( {k_3 \over \omega_3 } - {k_1 \over \omega_1 } \right)
\left( \epsilon_{zz} -1 \right)
\label{r51}
\end{equation}
For the back scattering
Eq. (\ref{r5}) gives
\begin{equation}
C\approx  \pm { 2 |q| k_2 \over \omega_{1,3} \omega_B  m} 
\left( \epsilon_{zz} -1 \right)
\label{r7}
\end{equation}

For the plasma waves on the mass surface $\epsilon_{zz} =0$ and
\begin{eqnarray}
&&
C = \pm { 2|q| k_2 \over  \omega_{1,3}  \omega_B  m}
\mbox{} \nonumber \\ \mbox{}
&&
u({\bf k}_3 ,{\bf  k_2},{\bf k}_1)  =
{ 2^2 q^2 k_2  ^2 ( 2 \pi)^5 \hbar   \over m^2 \omega_B^2 }
\omega_2 
\delta({\bf k}_3 -{\bf  k_2}-{\bf k}_1) \, \delta(\omega_3 - \omega_2 - \omega_1)  
\label{r71}
\end{eqnarray}

We can compare this probability with the unmagnetized case (\cite{Tsytovich}).
In the strongly magnetized plasma the probability of Raman scattering is decreased by
a ratio $\omega_{1,3} ^2 /\omega_B^2$. This is similar to the suppression of the
Thompson scattering, whose probability is decreased by the same ratio.

It is convenient to describe the wave distribution in terms
of photon occupation numbers
\begin{equation}
n_{\bf k}  ={ E^2 _{\bf k}  \over \hbar \omega ( {\bf k} ) }
\label{kk8}
\end{equation}
where $ E^2_{\bf k} $ is the spectral energy density, which is related
to the total energy density
\begin{equation}
W = \int { d {\bf k}  \over ( 2  \pi)^3} E^2_{\bf k} 
\label{kk9}
\end{equation}

The kinetic equation for the Raman scattering are (\cite{Melrosebook1},
Eq. (5.93))
\begin{eqnarray}
&&
{d n_1 ({\bf k}_1 )\over d t}
=- \,
  \int { d {\bf k}_3  \over ( 2 \pi) ^3}
 \int { d {\bf k}_2  \over ( 2 \pi) ^3}
u({\bf k}_3 , {\bf k}_1 ,{\bf k}_2 )
 \left(  n_1  ({\bf k}_1 )  n_2  ( {\bf k}_2  ) -
n_3 ( {\bf k}  ) \left(   n_1  ({\bf k}_1 ) + n_2  ( {\bf k}_2  ) \right) 
 \right)
\mbox{}  \nonumber \\ \mbox{} &&
\hskip 1 truein  +  \Gamma_3 n_3 ( {\bf k}_3  ) ,
\mbox{} \label{kd90} \\ \mbox{} &&
{d n_2 ( {\bf k}_2  )\over d t}
=- \,  \int { d ({\bf k}_1 ) \over ( 2 \pi) ^3}
 \int { d {\bf k}_3  \over ( 2 \pi) ^3}
u({\bf k}_3 , {\bf k}_1 ,{\bf k}_2 )
\left(  n_1  ({\bf k}_1 )  n_2  ( {\bf k}_2  ) -
n_3 ( {\bf k}_3  ) \left(   n_1  ({\bf k}_1 ) + n_2  ( {\bf k}_2  ) \right)
 \right)  
\mbox{}  \nonumber \\ \mbox{} &&
\hskip 1 truein +    \Gamma_2  n_2  ( {\bf k}_2  ),
\mbox{} \label{kd91} \\ \mbox{} &&
{d n_  3 ({\bf k}_3  )\over d t}
=\,  \int { d ({\bf k}_1 )\over ( 2 \pi)^ 3}
 \int { d {\bf k}_2  \over ( 2 \pi)^ 3}
u({\bf k}_3 , {\bf k}_1 ,{\bf k}_2 )
\left(  n_1  ({\bf k}_1 )  n_2  ( {\bf k}_2  ) -
n_3 ( {\bf k}_3  ) \left(   n_1  ({\bf k}_1 ) + n_2  ( {\bf k}_2  ) \right)
 \right)  
\mbox{}  \nonumber \\ \mbox{} &&
\hskip 1 truein  +    \Gamma_3  n_3 ( {\bf k}_3  )
 \label{kd92}
\end{eqnarray}

Where $ \Gamma_{1,2,3}$ are the corresponding damping or growth rates.
The most unimportant damping is the Landau damping of the 
plasma waves $ n_2 ( {\bf k}_2  )$ on the particles of plasma.
In (\ref{kd92}) the total derivative on the left hand side is 
\begin{equation}
{d \over  dt}  = \,{ \partial \over  \partial t }+ { \partial \omega \over  \partial
{\bf k} } \, { \partial \over  \partial {\bf r}} -
 { \partial \omega \over  \partial
{\bf r}}  \, { \partial \over  \partial {\bf k}} 
\label{kd092}
\end{equation}
 with the  
second term describing the convective transport of the waves.

Next we integrate over transverse wave vectors to find the system of equations
describing the scattering of the waves propagating along magnetic field

\begin{eqnarray}
\hskip -.5 truein
&&
{d n_1 ({ k_1} )\over d t}
=- \,
  \int { d { k_ 2}  \over ( 2 \pi) }
 \int { d { k_3}  \over ( 2 \pi) }
u({ k_ 2} , { k_1} ,{ k_2} )
\left( n_1  ({ k_1} )  n_2  ( { k_2}  ) -
n_ 3 ( { k_3}  ) \left(   n_1  ({ k_1} ) + n_2  ( { k_2}  ) \right)
 \right) 
\mbox{} \nonumber \\ \mbox{} &&
 \hskip 1 truein +  \Gamma_ 1 n_ 1( { k_ 1}  ) ,
\mbox{} \label{kd901} \\ \mbox{} &&
{d n_2 ( { k_2}  )\over d t}
=- \,  \int { d { k_1} \over ( 2 \pi) }
 \int { d { k_ 3}  \over ( 2 \pi) }
u({ k_ 2} , { k_1} ,{ k_2} )
\left(  n_1  ({ k_1} )  n_2  ( { k_2}  ) -
n_ 3( { k_ 3}  ) \left(   n_1  ({ k_1} ) + n_2  ( { k_2}  ) \right)
 \right) \mbox{} \nonumber \\ \mbox{} &&
 \hskip 1 truein  +    \Gamma_2  n_2  ( { k_2}  ),
\mbox{} \label{kd911} \\ \mbox{} &&
{d n_   3 ({ k_ 3}  )\over d t}
=\,  \int { d { k_1} \over ( 2 \pi)}
 \int { d { k_2}  \over ( 2 \pi)}
u({ k_ 2} , { k_1} ,{ k_2} )
\left(  n_1  ({ k_1} )  n_2  ( { k_2}  ) -
n_ 3 ( { k_ 3}  ) \left(   n_1  ({ k_1} ) + n_2  ( { k_2}  ) \right)
 \right) 
 \mbox{} \nonumber \\ \mbox{} &&
 \hskip 1 truein  +    \Gamma_ 3  n_ 3 ( { k_ 3}  )
 \label{kd919}
\end{eqnarray}
where
\begin{equation}
n_i (k_i) =  \int { d ^2 k_{i, \perp} \over ( 2 \pi) ^2} n_i ({\bf k}_i),
\hskip .2 truein i=1,2,3
\label{kd921}
\end{equation}
and 
\begin{equation}
u({ k_ 2} , { k_1} ,{ k_2} )
=
{ 2^2 q^2 k_2  ^2 ( 2 \pi)^5 \hbar \omega_2  \over m^2 \omega_B^2 }
\delta({ k_3} -{ k_2}-{k_1}) \, \delta(\omega_3 - \omega_2 - \omega_1)  
\label{r712}
\end{equation}

The two $\delta$-functions in $u({ k_ 2} , { k_1} ,{ k_2} )
$ allow us to perform integrations in (\ref{kd919}). 

\begin{eqnarray}
&&
{d n_1 ({ k_1} )\over d t}
=- \, { 2^2 q^2 k_2  ^2 ( 2 \pi)^3 \hbar \omega_2  \over m^2 \omega_B^2 
\left( {\partial \omega_3  \over \partial k_3 } \right) }
\left( n_1  ({ k_1} )  n_2  ( { k_2}  ) -
n_ 3 ( { k_3}  ) \left(   n_1  ({ k_1} ) + n_2  ( { k_2}  ) \right)
 \right) +  \Gamma_ 1 n_ 1 ( { k_ 1}  ) ,
\mbox{} \label{kd902} \\ \mbox{} &&
{d n_2 ( { k_2}  )\over d t}
=- \,   \, { 2^2 q^2 k_2  ^2 ( 2 \pi)^3 \hbar \omega_2  \over m^2 \omega_B^2 
\left( {\partial \omega_1  \over \partial k_1} \right) }
\left(  n_1  ({ k_1} )  n_2  ( { k_2}  ) -
n_ 3 ( { k_ 3}  ) \left(   n_1  ({ k_1} ) + n_2  ( { k_2}  ) \right)
 \right) +    \Gamma_2  n_2  ( { k_2}  ),
\mbox{} \label{kd912} \\ \mbox{} &&
{d n_   3({ k_ 3}  )\over d t}
=\,   \, { 2^2 q^2 k_2  ^2 ( 2 \pi)^3 \hbar \omega_2  \over m^2 \omega_B^2 
\left( {\partial \omega_2  \over \partial k_2 } \right) }
\left(  n_1  ({ k_1} )  n_2  ( { 
k_2}  ) -
n_ 3( { k_3}  ) \left(   n_1  ({ k_1} ) + n_2  ( { k_2}  ) \right)
 \right)  +    \Gamma_ 3 n_ 3( { k_ 3}  )
 \label{kd9192}
\end{eqnarray}

%dimension:
%\begin{equation}
 %[ { 2^2 q^2 k_2  ^2 ( 2 \pi)^3 \hbar \omega_2  \over m^2 \omega_B^2
%\left( {\partial \omega_3  \over \partial k_3 } \right) }
%\ = {cm ^2 \over s}
%\end{equation}

Implicit  assumption in Eqs. (\ref{kd902}) - (\ref{kd9192}) 
 is  that the wave vectors on the right hand side of Eqs. 
(\ref{kd902}),  (\ref{kd912})  and (\ref{kd9192}) 
 satisfy the resonance conditions.
Thus in Eq. (\ref{kd902}), for example, the two independent variables are time
 $t$ and wave vector $k_1$. The  explicit relations between wave vectors,
corresponding to Eqs. (\ref{kd902}),  (\ref{kd912}) and  (\ref{kd9192}), 
are 
\begin{eqnarray}
&& 
k_3\approx k_2  =  2 T^2 |k_1| + k_o ,
\hskip .2 truein 
\omega_2 =  ( 2 T^2 |k_1| + k_o - 2 |k_1| ) c 
 \mbox{} \nonumber \\ \mbox{}  &&
|k_1| = { k_2 - k_o \over 2 T^2  },
\hskip .2 truein 
k_3\approx k_2 ,
\hskip .2 truein 
\omega_2 = ( k_2 - { k_2 -k_o \over T^2} ) c
\mbox{} \nonumber \\ \mbox{}  &&
|k_1| \approx  { k_3- k_o \over 2 T^2  },
\hskip .2 truein 
k_2 \approx k_3,
\hskip .2 truein 
\omega_2 = ( k_3 - { k_3 -k_o \over T^2} ) c
\label{ttt1}
\end{eqnarray}
respectively for Eqs. (\ref{kd902}),  (\ref{kd912})  and (\ref{kd9192}).
The quantity 
 $ k_o = \sqrt{2 T} \omega_p$ is the cross-over wave vector, where 
the dispersion relation for the plasma waves intersect the vacuum dispersion
relation.

From Eqs. (\ref{ttt1}) and (\ref{n6}) we find that the induced Raman scattering
occurs in the frequency region
\begin{equation}
 \sqrt{2 T} \omega_p \leq kc \leq  2 \sqrt{2T} \omega_p 
\label{ttt2}
\end{equation}
This frequency window (in the plasma frame) corresponds
to the frequencies in the pulsar frame in the range
$ 2 \sqrt{ T \lambda \omega_B  \Omega /\gamma_p } 
\leq kc \leq  4 \sqrt{ T \lambda \omega_B  \Omega /\gamma_p }$, which is
$ 5 \times 10^8 {\rm rad \, sec^{-1}}\leq kc \leq  10^9 
{\rm rad \, sec^{-1}}$ which falls into the observed frequency range.

%Note: we neglected a difference in $k_2$  and $k_3$  in \ref{ttt1}

\section{ Application to Pulsars}
\label{Application}

In this section we consider the possible applications 
of the induced Raman scattering to the wave propagation and 
spectrum formation in the pulsar magnetospheres.

\subsection{ Scattering of the Pump Wave}
\label{ScateringPumpWave}

There is threshold intensity of the pump wave
that defines what is known as 
 strong induced Raman scattering.
The rate of induces plasmon production, which
depends on the intensity of the pump wave, should exceed the
rate of loss due to the various damping processes.
The end result of such strong  scattering  is a total depletion
of the initial pump wave. 
The  threshold  photon density for strong  Raman scattering
 is determined from (\ref{kd912})
\begin{equation}
  \int  { d  k_1  \over  2 \pi}
 \int { d k_3  \over  2 \pi}
u(k_3 ,  k_1 , k_2 )
n_3 (  k_3  ) =  \Gamma_2
 \label{kd93}
\end{equation}
where $\Gamma_2$ is the damping rate of plasma waves.

Linear 
Landau damping rate in the relativistic plasma ($T_p \gg 1$) is 
(\cite{LyutikovPhD}
\begin{equation}
\Gamma _2 = { \pi \omega_p^4  \over T k c  \omega^2}
\left( \gamma^3 { \partial f \over \partial \gamma } \right)_{res}
 \label{kd94}
\end{equation}
with 
\begin{equation}
 \gamma_{res} ^2 = { k_0 T_p^2 \over \Delta k},
\hskip .2 truein 
k_0  c = \omega_0 = \sqrt{2  T_p} \omega_p,
\hskip .2 truein
 \Delta k = k - k_0
 \label{kd97}
\end{equation}
The equations are valid for $  \Delta k \leq k_0$.

With a power law distribution function
$  f \propto \gamma^{-\alpha} , \, \alpha> 2 $
we find 
\begin{equation}
\Gamma _2 \approx { \pi \omega_p \over T_p^{3/2} } 
\left(  {   k_0 T_p^2 \over \Delta k} \right)^{ -\alpha/2+1 } 
 \label{kd98}
\end{equation}
 Note that as $  \Delta k \rightarrow 0$, i.e. when the 
wave vector approaches the cross-over point where the phase speed is
equal to the speed of light, $\Gamma _2 \rightarrow 0$ since there no particles
with the speed  equal to the speed of light.

The threshold spectrum of the incoming radiation is then given by
\begin{equation}
n_3 \approx { m ^2 c^3 T^{ - { 2+ 5 \alpha \over 4}} \over 64 \pi^2 \hbar q^2 } 
{\omega_B^2 \over \omega_p^2} 
\left( { \Delta k c \over \omega_p } \right)^{-1+ \alpha/2}
\label{kd981}
\end{equation}
Estimating $ \Delta k \approx k_o$ we find the characteristic threshold
photon density:
\begin{equation}
n_3^{th} \approx 10^{-2} { m^2 c^3 \omega_B^2 \over q^2  \hbar \omega_p^2} \,
T^{ -1-\alpha} = 2.7 \times 10^{22} \, \mbox{ ${1\over cm^2}$}
\,  B_{12}\, \lambda_5^{-1} \, P_{0.2} \, T^{-3}_{10}\,
R_9 ^{-3} \gamma_{p,10}
\label{kd9810}
\end{equation}
where $B_{12} = B/ 10^{12} {\rm G}\, $ is a magnetic field at the 
surface of the neutron star, $\lambda_5 =  \lambda/10^5 \,$ is the multiplicity factor,
i.e. the ratio of the plasma density to the Goldreich-Julian density $n_{GJ}= 
{\bf \Omega \cdot B}/( 2\,
\pi\, e\, c)$, $
P_{0.2} =  P/0.2 {\rm sec} $ is the pulsar period, $ T_{10} = T/10\, $ is the
relativistic invariant temperature, $R_9 = R/10^9 {\rm cm}\,$ is the 
distance from the neutron star, $
\gamma_{p,10} = \gamma_{p}/10$  is the average streaming Lorentz factor 
of the secondary plasma and we used
 $\alpha =2$. Note, that this is an {\it upper} bound on the
threshold photon density (for $ k \approx 2 k_o$).
At the lower frequencies $  k \approx  k_o$ the threshold
values are much lower.

For the relativistic 
Gaussian-type distribution $ f \propto 1/T \times \exp\{ -\gamma/T \}$
the damping rate is 
\begin{equation}
\Gamma _2 \approx \left( { \omega_p \over \Delta k c } \right) ^{3/2} 
{\omega_p \over T^{3/4}} \exp \left\{ - \left( {2  \omega_p^2 T \over
\Delta k^2\, c^2 } \right)^{1/4} \right\}
 \label{kd99}
\end{equation}
(compare with \cite{LominadzeMikhailovskii} Eq. (2.10)).
The spectrum of the incoming radiation is then given by
\begin{equation}
n_3 \approx { m ^2 c^3 \omega_B^2 \over 128 \pi^2 q^2 \hbar  T^{9/4} \omega_p^2}
\left( { \Delta k c \over \omega_p} \right) ^{-3/2}
\exp \left\{ - \left( {2  \omega_p^2 T \over
\Delta k^2\, c^2 } \right)^{1/4} \right\}
\label{kd991}
\end{equation}
The characteristic threshold
photon density is in this case
\begin{equation}
n_3^{th} \approx 10^{-4} { m^2 c^3 \omega_B^2 \over q^2 T^3 \hbar \omega_p^2}=
 5 \times 10^{-5} { m^2 c^3 \omega_B \gamma_p
\over q^2 T^3 \hbar \Omega \lambda} =
2.7 \times 10^{20} \, \mbox{ ${1\over cm^2}$}
\,  B_{12}\, \lambda_5^{-1} \, P_{0.2} \, T^{-3}_{10}\,
R_9 ^{-3} \gamma_{p,10} 
\label{kd9811}
\end{equation}
(upper bound)
where we used $ \omega_p^2 =  2 \omega_B \Omega \lambda/\gamma_p$.

For the comparable values of the temperature, i.e. for the 
similar dispersion of the energies of the plasma particles, the threshold
in the case of a power law distribution is much larger than in the case
of Maxwell-type distribution. 
This is due to the larger damping rate in the case of 
a power law distribution since there are more particles satisfying the
Cherenkov resonance in this case than in the case of Maxwell-type distribution.

The estimates (\ref{kd9810})  and (\ref{kd9811}) are done
for the one dimensional (in ${\bf k} $) photon density in the plasma frame.
The photon densities (\ref{kd9810}) and (\ref{kd9811})
are the one dimensional photon densities.  

The energy flux per unit interval of frequencies is given in terms of
photon density as follows:
\begin{equation}
F(\nu)  d \nu = { n({\bf k},{\bf r} ) c \hbar \omega d {\bf k}  \over 
8 \pi^3 } = { n(k,{\bf r} ) c \hbar \omega d k \over 2 \pi} 
 = { n(k,{\bf r} )  \hbar \omega  d \nu}
\label{kd9001}
\end{equation}
Where we took into account that the dispersion relation
of the electromagnetic waves is almost vacuum like, so $ k\approx \omega/c $.

We recall next, that both three dimensional occupation 
numbers $ n({\bf k},{\bf r} )$ and one dimensional  photon density $n(k,{\bf r} )$
(Eq. (ref{kd921})) are relativistic invariants. 
It follows from (\ref{kd9001}) that 
\begin{equation}
{ F(\nu)  \over \nu } = \mbox {inv}
\label{kd9002}
\end{equation}
(for the one dimensional flux, i.e. integrated over transverse wave vectors).
 Thus, the flux in the observer frame
is
\begin{equation}
F^{\prime} (\nu ^{\prime}) = 
{  2 \pi n(k^{\prime} ,{\bf r} )  \hbar \nu^{\prime}  }
\mbox{ in ${\rm { erg \over cm^2 \, sec \, Hz} } $}
\label{kd903}
\end{equation}

We can estimate the observed flux (in Jankys) at the Earth:
\begin{equation}
F^{{\rm obs}} (\nu^{\prime} )=  F^{\prime} (\nu ^{\prime}) 
\left( {R \over d}\right)^2\, 
\label{r41}
\end{equation}
 Using Eqn. (\ref{kd9810}), (\ref{kd9811}), (\ref{kd903}) and (\ref{r41}) we
find
\begin{equation}
F^{{\rm obs}}  = 
\left\{ \begin{array}{ll}
\phantom{ {{{ {a\over b} \over {a\over b} } } \over {{ {a\over b} \over {a
\over b} } } } }
{10^{-2} \over T^{1+ \alpha} }  &\\
\phantom{ {{{ {a\over b} \over {a\over b} } } \over {{ {a\over b} \over {a
\over b} } } } }
{10^{-4} \over T^3 } &
\end{array} \right.
\times { \pi m^2 c^3 \omega_B^{\ast}  \gamma_p  \nu \over 
q^2 \Omega \lambda } \left( {R_{NS} \over d}\right)^2
\left( {R _{NS} \over R }\right) 
\label{r4112}
\end{equation}
 for the power law and exponential distributions correspondingly.

Numerical estimates give
\begin{equation}
F^{{\rm obs}}  =
\left( \begin{array}{ll}
30 &\\
3  \times 10^{-1} &
\end{array} \right) \mbox{Jy}
\label{r4113}
\end{equation}
These are comparable with the fluxes observed from 
bright pulsars. Since the estimated threshold intensities
are upper bounds, we conclude that stimulated Raman
scattering 
may be an important factor in the formation of pulsar radio
emission.
Estimates for the millisecond pulsars give
about an order of magnitude lower values.

\subsection{Saturation of Cyclotron Instability by the
Induced Raman Scattering}
\label{Saturation}

Cyclotron-Cherenkov instability, operating in the outer parts of magnetosphere
at the anomalous
Doppler resonance
\begin{equation}
\, \omega({\bf k}) - k_{\parallel} v_{\parallel} -s
 {\, \omega_B \over  \gamma }
=0 \, \, \mbox{ for $s< 0$}
\label{ba}
\end{equation} 
has been suggested as a possible mechanism of pulsar radio
emission (\cite{KawamuraSuzuki}, \cite{LominadzeMachabeliMkhailovsky},
\cite{Kazbegi}, \cite{LyutikovPhD}). In this section 
we consider the nonlinear saturation of the cyclotron-Cherenkov instability
by the induced Raman Scattering. 
To simplify the consideration, 
we assumes that the 
product electromagnetic waves (labeled by index $1$) leave the
region of resonant interaction, so we can assume that their
photon density is $n_1(k_1) \approx 0$.
We are interested in the time asymptotic limit for the 
intensities of the transverse waves, excited by some
instability mechanism with a growth rate $\Gamma_3$. 
Neglecting the wave convection we then find from 
Eqs. (\ref{kd902}) and (\ref{kd9192}):

\begin{eqnarray}
&&
 \,   \, { 2^2 q^2 k_2  ^2 ( 2 \pi)^3 \hbar \omega_2  \over m^2 \omega_B^2 
\left( {\partial \omega_1  \over \partial k_1} \right) }
n_ 3  ( k_ 3) 
+    \Gamma_2  n_2  ( { k_2}  )=0,
\mbox{} \label{kd9121} \\ \mbox{} &&
 - \, { 2^2 q^2 k_2  ^2 ( 2 \pi)^3 \hbar \omega_2  \over m^2 \omega_B^2 
\left( {\partial \omega_2  \over \partial k_2 } \right) }
 n_2  ( { k_2}  )
 +   \Gamma_ 3 n_ 3( { k_ 3}  )=0
 \label{kd91921}
\end{eqnarray}

Comparing 
 Eqs. (\ref{kd93}) and (\ref{kd9121}) we find that they are identical:
the threshold intensity for the strong
Raman scattering is equal to the saturation intensity. This is not a
surprising fact, since in both cases it is the damping of the
plasma waves that controls the electromagnetic wave intensity.
In Section \ref{ScateringPumpWave} we found that the
corresponding intensities are comparable to the observed ones,
so that the nonlinear saturation by the induced Raman scattering
may provide an observed emissivities.

We also note in 
this context, that at a given radius, the induced Raman scattering
operates in a limited frequency range (\ref{n6}).
So, if the instability that produces radio emission is 
intrinsically broad band (with the bandwidth of the growing
mode larger than the range in which the induced Raman scattering
operates), then the effects of the induced Raman scattering
would produce a dip in the spectrum, corresponding to that
frequency range. 
If, on the other hand, the instability that produces radio emission is 
intrinsically  narrow band with the central frequency of the instability
falling into the operational range of the induced Raman scattering and 
changing with radius to mimic the observed broadband spectrum,
then the induced Raman scattering can be an effective 
saturation mechanism for a larger range of frequencies. 

\section{Conclusion}
\label{Conclus}

In this work we considered  
 induced Raman scattering 
of a transverse  electromagnetic wave propagating along magnetic field in 
a pair plasma.
We found 
 that induced Raman scattering
can be an important factor for the
wave propagation in the pulsar magnetosphere and as a nonlinear saturation 
mechanism for the electromagnetic instabilities ( cyclotron-Cherenkov 
instability, for example). The threshold intensities
for strong Raman scattering, which are determined by the damping
rate of the Langmuir waves in the relativistic plasma, are comparable
to the observed fluxes from radio pulsars. 

One of the major limitations of this work is the assumption of the
propagation along the magnetic field. The natural extension of this work to
oblique propagation would involve several complications:
(i) there will be a distinction in the transition probabilities
 between extraordinary mode (with the 
electric field of the waves perpendicular to the ${\bf k} - {\bf B}$ plane)
and ordinary mode (with the 
electric field  in the  ${\bf k} - {\bf B}$ plane), 
(ii)  ordinary mode has an sensitive dependence on the angle of propagation, 
so that for the  ordinary mode  the Raman scattering will occur only in a
very limited angle around magnetic field; the scattered waves will be a backward
propagating,
(iii)  for the extraordinary mode  the Raman scattering is possible for 
large angles of propagation; in this case both electromagnetic waves can be forward
propagating.

\acknowledgements
I would like to thank Roger Blandford, Peter Goldreich  and Gia Machabeli
for their useful 
comments.

{}

\newpage

FIGURE CAPTIONS.

Fig. 1.
%ref{RamanFig}.
Kinematics of backwards Raman scattering in the plasma frame.
The vacuum dispersion relation $\omega= k c$, the transverse wave
dispersion relations $\omega= k c (1-\delta)$ and the plasma wave
dispersion relations (which starts from
$ \left( {2\over T} \right)^{1/2} \omega_p$) for the parallel
propagation are shown. The bold faced vectors represent ${\bf k}
\equiv \{ \omega, k \}$.
Cross-over point is given by $k_o$ and $\omega_o$.
Only waves with $ k > k_t $ can be scattered.

\newpage

%setcounter{section}{0}
%newcounter{label}
%setcounter{label}{0}
%setcounter{figure}{0}

\vskip 2 truein 

\begin{figure}
\psfig{file=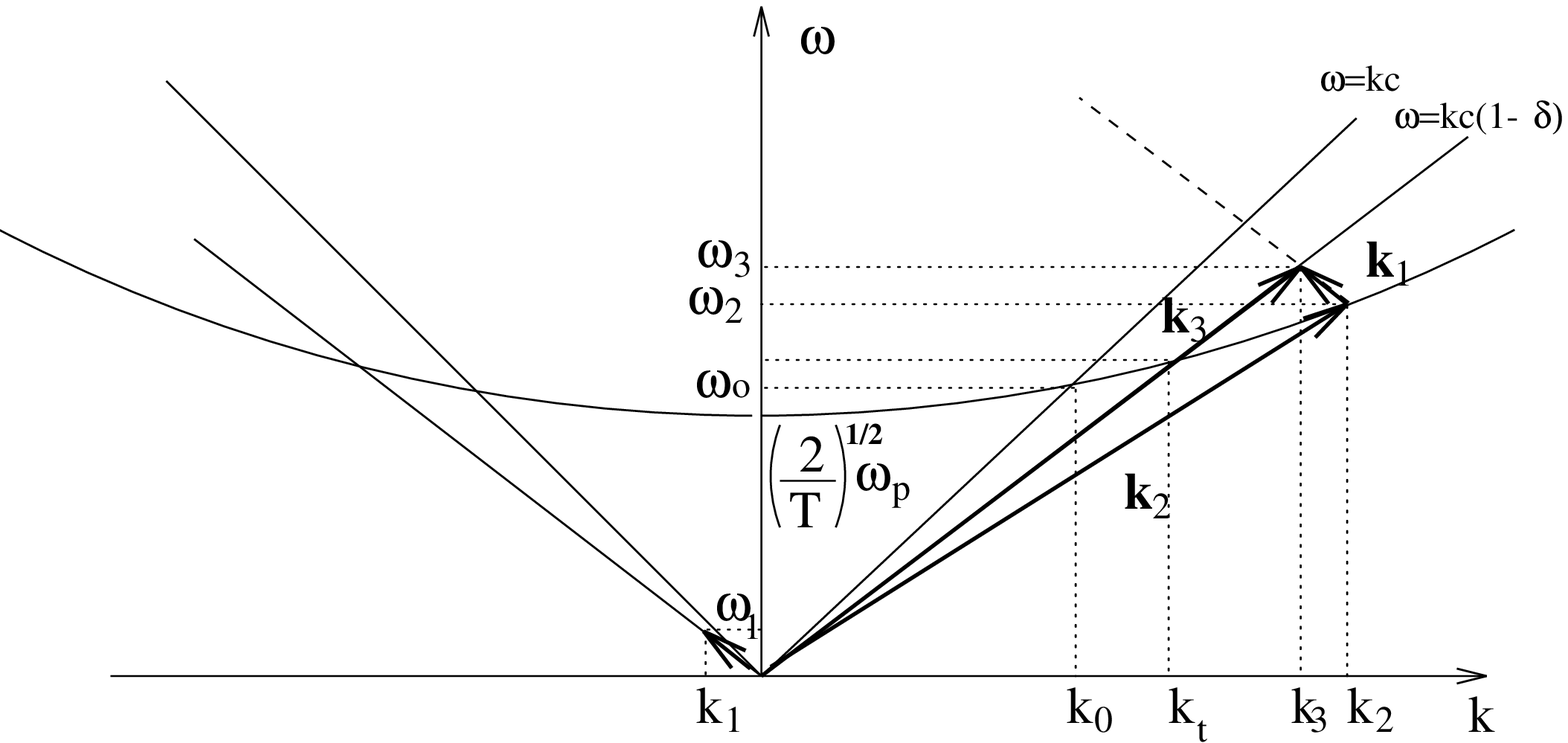,width=15.0cm}
\label{RamanFig}
\end{figure}

\end{document}